\documentclass[aps,prb,twocolumn,groupedaddress,showpacs]{revtex4-1}
\usepackage[english]{babel}
\usepackage{stmaryrd}
\usepackage{amssymb}
\usepackage{amsfonts}
\usepackage{amsmath}
\usepackage{color}
\usepackage{xspace}
\usepackage{graphicx}
\usepackage{bm}

\begin{document}

\title{Static and dynamical spin correlations of the $S=$1/2 random-bond antiferromagnetic Heisenberg model on the triangular and the kagome lattices}
\author{Tokuro Shimokawa, Ken Watanabe, and Hikaru Kawamura}
\affiliation{Department of Earth and Space Science, Graduate School of Science, Osaka University, Toyonaka, Osaka 560-0043, Japan}

\date{\today}

\begin{abstract}
Inspired by the recent theoretical suggestion that the random-bond $S=1/2$ antiferromagnetic Heisenberg model on the triangular and the kagome lattices might exhibit a randomness-induced quantum spin liquid (QSL) behavior when the strength of the randomness exceeds a critical value, and that this ``random-singlet state'' might be relevant to the QSL behaviors experimentally observed in triangular organic salts $\kappa {\rm -(ET)_2 Cu_2 (CN)_3}$and ${\rm EtMe_3 Sb[Pd(dmit)_2]_2}$ and in kagome herbertsmithite ${\rm ZnCu_3(OH)_6Cl_2}$, we further investigate the nature of the static and the dynamical spin correlations of these models. We compute the static and the dynamical spin structure factors, $S({\bf q})$ and $S({\bf q},\omega)$, by means of an exact diagonalization method. In both triangular and kagome models, the computed $S({\bf q},\omega)$ in the random-singlet state depends on the wavevector ${\bf q}$ only weakly, robustly exhibiting gapless behaviors accompnied by the broad distribution extending to higher energy $\omega$. Especially in the strongly random kagome model, $S({\bf q},\omega)$ hardly depends on ${\bf q}$, and exhibits an almost flat distribution for a wide range of $\omega$, together with a $\omega=0$ peak. These features agree semi-quantitatively with the recent neutron-scattering data on a single-crystal herbertsmithite. Furthermore, the computed magnetization curve agrees almost quantitatively with the experimental one recently measured on a single-crystal herbertsmithite. These results suggest that the QSL state observed in herbersmithite might indeed be the randomness-induced QSL state, {\it i.e.\/}, the random-singlet state.
\end{abstract}

\maketitle

\section{\label{sec:INTRODUCTION}Introduction}

 Geometrically frustrated magnets have attracted a long-standing and ongoing attention in the field of condensed matter physics because they often give rise to a variety of nontrivial thermodynamic states. In particular, the quantum spin liquid (QSL) state having no magnetic long-range order (LRO) has been extensively investigated both theoretically and experimentally. Anderson proposed the resonating valence bond (RVB) state as a possbile ground state of the $S=1/2$ Heisenberg antiferromagnet on the triangular lattice\cite{Anderson}. According to many subsequent theoretical studies, however, it is now widely believed that the ground state of a simple $S=1/2$ antiferromagnetic (AF) Heisenberg model on the triangular lattice with the nearest-neighbor bilinear interaction exhibits a  N$\acute{\rm e}$el LRO with the three-sublattice $120^{\circ}$ structure\cite{Bernu, Capriotti, White}.

 By contrast, the ground state of the $S=1/2$ Heisenberg model on the kagome lattice, which has stronger frustration than the triangular lattice, is quite likely to be non-magnetic. Namely, the absence of the magnetic LRO in the $S=1/2$ kagome AF Heisenberg model has now been established from various numerical studies\cite{Elser,Singh,Leung,Lecheminant,Waldtmann,Misguich,Singh2,Jiang,Lauchli2,Sindzingre,Nakano,Lauchli,Yan,Depenbrock,Sugiura,Nishimoto,Nakano2}, although the precise nature of its ground state still remains controversial. Various candidates, including the gapped $\mathbb{Z}_2$ spin-liquid \cite{Sachdev, Lu, Yan, Depenbrock}, the gapless U(1) spin liquid\cite{Hastings, Ran, Nakano, Iqbal, Iqbal2}, the chiral spin liquid\cite{Messio} and the valence bond crystal\cite{Marston, Singh2, Evenbly} $etc$, have been proposed.

 Along with these theoretical studies, experimental quest for the QSL has also been persued. As a result, several promising candidate materials were recently reported. In the triangular system, certain organic salts such as $\kappa {\rm -(ET)_2 Cu_2 (CN)_3}$\cite{Shimizu,Kurosaki,Shimizu2,Ohira,SYamashita, MYamashita,Manna,Jawad,Pratt,Poirier,Sedlmeier,Goto,Itoh} and ${\rm EtMe_3 Sb[Pd(dmit)_2]_2}$\cite{Itou, Itou2, MYamashita2, SYamashita2,DWatanabe,Jawad2} were reported to exhibit a QSL-like behavior, {\it i.e.\/}, exhibit no magnetic LRO down to a very low temperature. These organic materials show gapless (or nearly gapless) behaviors at low temperatures. For example, the specific heat\cite{SYamashita, SYamashita2} and the thermal conductivity\cite{MYamashita, MYamashita2} exhibit the behavior linear in the absolute temperature $T$. Obviously, in understanding the QSL-like properties of these triangular organic salts, some extension beyond the simplest nearest-neighbor Heisenberg model is needed\cite{LiMing, Morita, Motrunich, Lee, Yunoki, Lee2, Qi, Tocchio}.

 In the kagome system, herbertsmithite ${\rm ZnCu_3(OH)_6Cl_2}$ was reported to be a candidate of QSL, again showing gapless behaviors in various physical quantities\cite{Shores, Helton, Vries, Mendelse, Helton2,Freedman, Han, Han2}. The true nature of these gapless QSL candidates experimentally observed both in the triangular and the kagome lattice AFs has still remained obscure in spite of much theoretical and experimental efforts devoted to the issue.

 For years, major part of theories have tried to elucidate the experimentally observed QSL behaviors as properties of a clean and regular system. By contrast, it was proposed recently in Refs.\cite{Watanabe} and \cite{Kawamura} that the quenched bond-randomness, together with the geometrical frustration effect, might be essential in stabilizaing the experimentally observed gapless QSL state both in the triangular organic salts and in the kagome herbersmithite. Such a randomness-induced QSL state is called the ``random-singlet state''\cite{Dasgupta, Bhatt, Fisher, Lin} or the ``valence bond glass state''\cite{Tarzia, Singh3}, where the spin singlet is formed in a spatially random manner. Possible importance of the quenched site-randomness in kagome herbertsmithite was also pointed out in Ref.\cite{Singh3}.

 The origin, or even the existence of the quenched randomness in triangualr organic salts and kagome herbersmithite is a non-trivial matter, especially in view of the fact that the theory requires a considerable amount of randomness, not an infinitesimal one, to realize the QSL-like behavior \cite{Watanabe,Kawamura}.

 In the case of the triangular organic salts, it was suggested that the randomness for the spin degrees of freedom was self-generated at low temperatures via the random freezing of the electric-polarization degrees of freedom inherent to these organic salts consisting of molecular dimers \cite{Watanabe}. In fact, the measured ac dielectric constant of these organic salts exhibited a glassy response even on a macroscopic time scale of kHz in the temperature region where the QSL behavior was observed in the spin degree of freedom\cite{Jawad}.

 In the case of kagome herbertsmithite, the quenched randomness comes from the random substitution of non-magnetic Zn$^{2+}$ by magnetic Cu$^{2+}$, located on the triangular layer adjacent to the kagome layer \cite{Freedman,Kawamura}. Note that the kagome layers in herbertsmithite are separated by $\rm [Zn(OH)_6]^{4-}$ octahedral units whose Zn$^{2+}$ constitutes the triangular layer. It was reported that about 15$\%$ of Zn$^{2+}$ on the triangular layer was randomly substituted by Cu$^{2+}$, keeping the kagome layer intact\cite{Freedman}. Since Cu$^{2+}$ is a Jahn-Teller ion, such a random substituion would lead to a random Jahn-Teller distortion of the $\rm [Cu(OH)_6]^{4-}$ octahedra, leading to the random modification of the exchange path, and subsequently the exchange strength, connecting the Cu$^{2+}$ on the kagome layer \cite{Tanaka}.

 Indeed, the numerical results on a simplified random $S=1/2$ AF Heisenberg model on the triangular and the kagome lattices appear to reproduce many of experimentally observed features\cite{Watanabe, Kawamura} of various thermodynamic quantities, including the $T$-lienar  low-temperature specific heat\cite{SYamashita, SYamashita2, Helton}, the gapless magnetic susceptibility occasionally accompanying a Curie-like tail\cite{Helton2}, and the gapless temperature dependence of the NMR relaxation rate 1/$T_1$ \cite{Shimizu2, Itou2}.

 In view of this apparent success of the random model in reproducing the experimentally observed QSL-like behaviors, it would be desirable to further investigate the nature of the static and the dynamical spin correlations of the randomness-induced QSL state, the random singlet state. For this purpose, in the present paper we compute by means of an exact diagonalization technique the static and the dynamical spin structure factors which are experimentally accessible via, {\it e.g.\/}, the elastic and the inelastic neutron scattering measurements. Such a comparison between theory and experiment might give further information in examining the validity of the randomness-induced QSL picture of the experimentally observed QSL states. In order to clarify the effect of frustration, we perform comparative calculations also on the random-bond $S=1/2$ AF Heisenberg model on the square lattice.

  Our numerical results corroborates the previous observation that both the triangular and the kagome models exhibit the randomness-induced QSL-like behavior when the randomness exceeds a critical value \cite{Watanabe,Kawamura}. Meanwhile, the unfrustrated square model persistently exhibits the AF LRO up to the maximal randomness without showing the QSL-like behavior. The result hilights an important role of frustration, along with the randomness and the quantum fluctuation, in stabilizing the random-singlet state.

 The random-singlet states in the triangular and in the kagome models have some mutual similarities, but also some differences. In the triangular case, the random-singlet state keeps a certain amount of AF short-range order even at the maximal randomness. While the dynamical structure factor $S({\bf q},\omega)$ of the triangular model exhibits a signature of the AF LRO and the magnon excitation in the regular and weakly random cases, it exhibits, in the strongly random case corresponding to the random-singlet state, a gapless behavir accompanied by a broad $\omega$-distribution extending to higher energy, which is dependen on the wavevector ${\bf q}$ only weakly. In the kagome case, by contrast, a signature of the AF LRO or the magnon excitation is hardly discernible either in the regular or in the random case. Peaky features of $S({\bf q},\omega)$ still retained in the regular case give way to gapless behaviors in the strongly random case corresponding to the random singlet state. Such features of $S({\bf q},\omega)$ are accompanied by an almost flat distribution in a wide range of $\omega$ and by an $\omega=0$ peak, which hardly dependes on the wavevector ${\bf q}$. Indeed, these features of the computed  $S({\bf q},\omega)$ of the strongly random kagome model are compared quite faborably with the recent inelastic-neutron scattering data on a single-crystal kagome herbersmithite \cite{Han}.

 This paper is organized as follows. In Sec.~II, we present our model and the details of our numerical calculation. In Sec.~III and IV, we show the results of our numerical calculations in the case of the triangular and the kagome models, respectively. Section V is devoted to summary and discussion. For comparison, we also present the corresponding results for the unfrustrated square-lattice model in Appendix A. The detailed information about the shape of finite-size lattices employed in our exact-diagonalization calculation is given in Appendix B.

\section{The model and the method}

 Our model is the AF random-bond $S=1/2$ quantum Heisenberg model on the triangular and the kagome lattices, whose Hamiltonian is given by

\begin{eqnarray}
\mathcal{H}=\sum_{i,j} J_{i,j} {\bf S}_i \cdot {\bf S}_j
\end{eqnarray}
where ${\bf S}_i=(S_i^x,S_i^y,S_i^z)$ is a spin-1/2 operator at the $i$-th site on the lattice, while $J_{i,j}>0$ is the random nearest-neighbor AF coupling obeying the bond-independent uniform distribution between [$(1-\Delta ) J$,$(1+ \Delta )J$] with the mean $J$. The parameter $\Delta$ represents the extent of the randomness: $\Delta=0$ and $\Delta=1$ cases correspond to the regular and maximally random cases, respectively.

 According to Refs.\cite{Watanabe} and \cite{Kawamura}, the random-singlet ground state is realized when the randomness is stronger than a critical value $\Delta_{\rm c}$. In the triangular model, $\Delta_{\rm c}$ is estimated to be $\simeq 0.6$ where $\Delta_{\rm c}$ separates the AF phase and the random-singlet phase\cite{Watanabe}, while in the kagome model it is estimated to be $\Delta_{\rm c} \simeq 0.4$ where $\Delta_{\rm c}$ separates the  the randomness-irrelevant QSL phase ({\it e.g.\/}, the $Z_2$ spin-liquid phase) and the randomness-relevant random-singlet phase \cite{Kawamura}.

 In the present paper, we employ the exact diagonalization (ED) method in computing various physical quantities. The ED method is precise and is applicable even to systems with frustration, while it has a disadvantage of being limited to very small system sizes. In our computation, the total number of spins $N$ is $N=12, 18, 24, 30$ for $T=0$ and $N=12, 18$ for $T>0$, periodic boundary conditions being employed. Sample average is taken over 100 ($N=12,18,24$) and 50 ($N=30$) independent bond realization in the $T=0$ calculation, while 100 ($N=12$) and 30 ($N=18$) in the $T>0$ calculation in both cases of the triangular and the kagome models. The shape of the lattice is illustrated in Fig.13 and Fig.14 in Appendix B for the triangular and the kagome lattices, respectively. In what follows, the energy and temperature are normalized in units of $J$.

\section{\label{sec:Triangular lattice}Results I: the triangular lattice}

\begin{figure}[t]
 \includegraphics[width=7cm, angle=0]{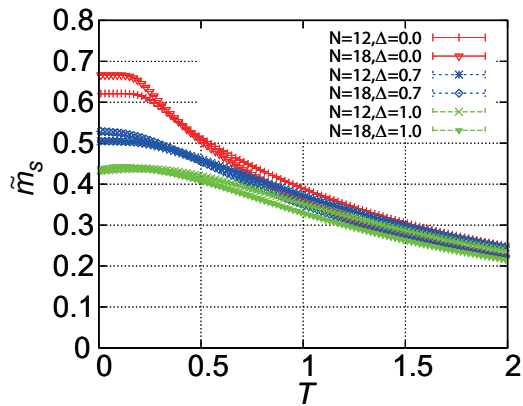}
 \caption{(Color online) The temperature dependence of the rescaled sublattice magnetization per spin $\tilde{m}_{\rm s}$ of the triangular-lattice Heisenberg antiferromagnet for the randomness $\Delta=0, 0.7$ and $1.0$ and for the sizes $N=12$ and 18. The definition of $\tilde{m}_{\rm s}$ is given in the text.}
 \label{}
\end{figure}

In this section, we study the ground-state and the finite-temperature properties of the random-bond $S=1/2$ AF Heisenberg model on the triangular lattice. We begin with the sublattice magnetization associated with the three-sublattice 120$^{\circ}$ structure. The sublattice magnetization per spin $m_{\rm s}$ may be defined by
\begin{eqnarray}
m_{\rm s}^2 &=& \frac{1}{3} \sum_{\alpha} \left[ \frac{1}{(N/6)(N/6+1)} \left( \sum_{i \in \alpha} {\bf S}_i\right)^2 \right] \nonumber \\
      &=& \frac{12}{N(N+6)} \sum_{\alpha} \left[ \sum_{i,j \in \alpha} \langle {\bf S}_i \cdot {\bf S}_j \rangle \right] 
\end{eqnarray}
where $\alpha=1,2,3$ denotes three triangular sublattices of the original triangular lattice, and the sum $i \in \alpha $ $(i,j \in \alpha)$ is taken over all site $i$ ($i$ and $j$) belonging to the $\alpha$-th sublattice. The symbol $\langle \cdots \rangle$ denotes the thermal average, and [$\cdots$] the average over the bond disorder. Note that $m_{\rm s}$ is normalized to give unity for the classical, perfectly ordered 120$^\circ$ structure.

 We also define the rescaled sublattice magnetization per spin $\tilde m_{\rm s}$  by subtracting the auto-correlated part as,
\begin{equation}
\tilde m_{\rm s}^2 = m_{\rm s}^2 - \frac{12N}{N(N+6)}\times \frac{3}{4} ,
\end{equation}
The subtraction of the auto-correlated part is made so that the sublattice magnetization tends to zero at $T \rightarrow \infty$ even for finite $N$. In the thermodynamic limit $N\rightarrow \infty$, $\tilde m_{\rm s}$ approaches $m_{\rm s}$ at any temperature.

 In Fig.~1, we show the temperature dependence of the rescaled sublattice magnetization per spin, $\tilde m_{\rm s}$, for $N=12$ and 18 for several values of the randomness $\Delta$. In the regular and weakly random cases,  $\tilde m_{\rm s}$ increases monotonically with decreasing the temperature, and eventually saturates. This increase of $m_{\rm s}$ at finite temperature is associated with the growth of the AF short-range order. When the randomness gets stronger, $\tilde m_{\rm s}$ is supressed as a result of the supression of the AF short-range order due to the randomness.

 An interesting observation is that, for the strongest randomness $\Delta=1$, $\tilde m_{\rm s}$ tends to {\it decrease\/} weakly with decreasing the temperature in the lower temperature range of $T\lesssim 0.2$, exhibiting some sort of crossover. This supression is weak, but tends to be more eminent for larger sizes. Such supression of the AF short-range order is compatible with the formation of the random-singlet-type state at $T\lesssim 0.2$. Similar supression was also reported in Ref.\cite{Watanabe} in the temperature dependence of $1/T_1$ in the similar temperature and size range. There, $Z_2$ vortex \cite{KawaMiya} was invoked to be a possible candidate of this crossover-like anomaly, since the supression was observed only for lattices larger than $N=18$. Further study is required to clarify the nature of this crossover phenomenon, which is observed only for modestly large systems with strong randomness.

\begin{figure}[t]
 \includegraphics[width=8cm]{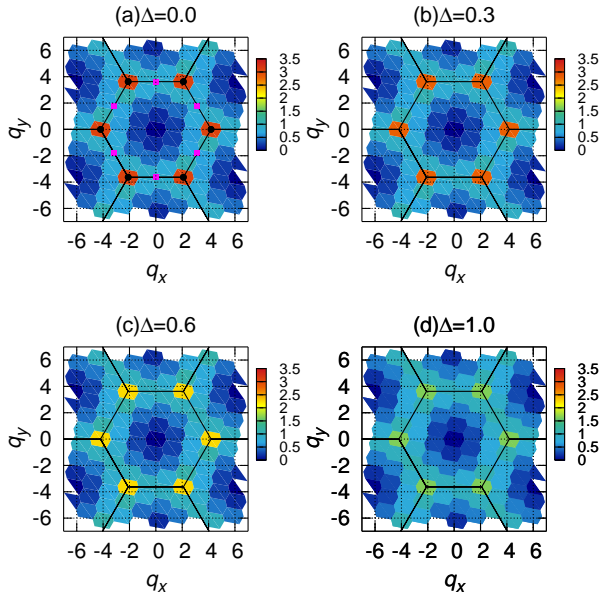}
 \caption{(Color online) Intensity plots of the static spin structure factors $S({\bf q})$ of the triangular-lattice Heisenberg antiferromagnet in the wavevector ($q_x,q_y$) plane for several values of the randomness $\Delta=0$ (a), 0.3 (b), 0.6 (c), and 1.0 (d). The lattice constant $a=1$ is the nearest-neighbor distance of the triangular lattice. The system size is $N=30$. The solid black line depicts the first Brillouin zone of triangular lattice. Each small hexagon corresponds to the resolution unit, and the center of each distorted hexagon is the wavevector point we can treat in $N=30$ system. The black point in Fig.(a) represents the K point, while the orange point the M point: see the text for details.
}
\end{figure}

 In order to get further information about the static spin correlations of the model, we investigate the static spin structure factor $S({\bf q})$ defined by
\begin{eqnarray}
S({\bf q})=\frac{1}{N} \left[ \langle |\sum_{j} {\bf S}_j e^{\mathrm{i} {\bf q} \cdot {\bf R}_j}|^2 \rangle \right],
\end{eqnarray}
where $\langle \cdots \rangle$ means the ground-state expectation, ${\bf q}$ is the wavevector, and ${\bf R}_j$ is the position vector at the site $j$. The computed static spin structure factor at $T=0$ is shown in  Fig.~2 as an intensity plot in the wavevector ($q_x,q_y$) plane for several values of the randomness, $\Delta=0$ (a), 0.3 (b), 0.6 (c), and 1.0 (d). The system size is $N=30$.

 Although the resolution is rather low due to the small system size, several characteristic features are clearly discernible from the figure. In the regular case $\Delta=0$, eminent peaks corresponding to the AF LRO are observed at the so-called K points, ${\bf q}=(q_x, q_y)=(\pm 2\pi/3, \pm2\pi/\sqrt{3})$ and $(\pm 4\pi/3, 0)$, where the length unit is taken as the nearest-neighbor distnace of the triangular lattice. As the randomness gets stronger, the peak at the K points is gradually suppressed. The AF LRO is expected to vanish for $\Delta>\Delta_{\rm c}\simeq 0.6$ \cite{Watanabe}. Such a phase-transition-like sharp change of behavior around $\Delta_{\rm c} \sim 0.6$, however, is not necessarily clear here, presumably due to the small system size of $N=30$. In fact, a rounded peak persists  at the K point even at the maximal randomness of $\Delta=1$, suggesting the persistance of the AF short-range order in the random-singlet state. Thus, in the triangular case, the random-singlet state coexists with the AF {\it short-range\/} order.

 In order to get further information about the dynamical spin correlations of the 
model, we next investigate the dynamical spin structure factor defined by
\begin{eqnarray}
S({\bf q}, \omega) &=& \sum_{n} \left[ |\langle \psi_{n}|  S_{q}^{z} | \psi_{0} \rangle|^2 \delta (\omega-(E_{n}- E_{0})) \right] , \nonumber \\
\end{eqnarray}
where $S_{\bf q}^z$ is the $z$-component of the Fourier transform of the spin operator, $\psi_n$ is the  eigenfunction of the Hamiltonian (1) whose eigenvalue is $E_n$, and $\psi_0$ is the ground-state eigenfunction with the eigenvalue $E_0$.  By using the continued fraction method\cite{Gagliano}, it may be rewritten as
\begin{eqnarray}
S({\bf q}, \omega) &=& - \lim_{\eta \to 0} \frac{1}{\pi} {\rm Im} \left[ \langle \psi_{0} | (S_{q}^{z})^{\dagger} \frac{1}{\omega +E_{0}-\mathcal{H} + i \eta} S_{q}^{z} |\psi_{0} \rangle \right] \nonumber \\
                   &=& - \lim_{\eta \to 0} \frac{1}{\pi} {\rm Im} \left[ \frac{\langle \psi_0 |(S_{q}^{z})^{\dagger} S_{q}^{z} | \psi_0  \rangle}{\omega + E_0 + i \eta -a_0-\frac{b_1^2}{z-a_1-\frac{b_2^2}{z-a_2-\cdots}}} \right],
\end{eqnarray}
where  $a_i$ and $b_{i+1}$ are the diagonal and sub-diagonal elements of the tri-diagonal Hamiltonian obtained by the Lanczos method. In implementing the continued-fraction expansion, we performed at least 1000 iterations. A small but finite $\eta$, $\eta=0.02$, is used.

 The $\omega$-dependence of $S({\bf q}, \omega)$ computed at the K point is shown in Fig.3, while the one at the M point, which is located at the midpoints of the 1st Brillouin zone (BZ) edges, {\it i.e.\/}, ${\bf q}=(0, \pm 2 \pi / \sqrt{3})$ and $(\pm \pi, \pm \pi / \sqrt{3})$, is in Fig.4. The randomness $\Delta$ is taken to be $\Delta=0$ (a), 0.3 (b), 0.6 (c), and 1.0 (d). Note the difference in the ordinate scale between Figs. 3 and 4, and between Figs.(a) and (b)-(d).
 
  Reflecting the AF short-range order of the triangular model, the $S({\bf q}, \omega)$ intensity tends to be larger at the K point (Fig.3) than at the M point (Fig.4) irrespective of the $\Delta$-value. The contrast between the K and the M points, however, tends to be milder for larger $\Delta$.

 In the regular case $\Delta=0$, the dominant peak observed in $S({\bf q},\omega)$ at the K point (Fig.3(a)) is a single magnon excitation. Indeed, the peak location in $\omega$ tends to zero when $N$ is increased toward the thermodynamic limit. Similar behavior is observed also in the weakly random case of $\Delta =0.3$: see the insets of Figs.3(a) and (b). As shown in Fig.4(a), several peaks are observed at the M point. According to the recent spin-wave analysis for the regular model, $S({\bf q}, \omega)$ at the M point exhibits a leading peak at $\omega\simeq 0.8$\cite{Mourigal}. Indeed, if one looks at our data of $\Delta=0$ for $N=24$ and 30, the dominant peak appears at a nearby position $\omega\simeq 1$.

 As the randomness becomes stronger getting into the random-singlet phase, $S({\bf q}, \omega)$ exhibit less peaky behavior both at the K and the M points.  As can be seen from Figs.4(c) and (d), it exhibits a broad distribution extending to higher $\omega$, with a finite intensity gradually growing at $\omega=0$. This demonstrates that the random-signlet state of the triangular model is indeed magnetically gapless both at the K and the M points. The observed gapless feature of $S({\bf q}, \omega)$ is fully consistent with the gapless behavior observed in several other observables in Ref.\cite{Watanabe}. While the intensity tends to be larger at the K point than at the M point, the difference tends to be smaller for larger $\Delta$ as in the case of the static spin structure factor.

 In the random-singlet state, $S({\bf q},\omega)$ exhibits a tail in $\omega$ in the higher-$\omega$ range. The asymptotic $\omega$-dependence of this tail is found to be exponential $\approx \exp [-\omega/\omega_0]$ with a characteristic energy scale $\omega_0$. Estimates of $\omega_0$ yields a value around 2$\sim $2.5.

\begin{figure}[t]
 \includegraphics[width=6.7cm,angle=270]{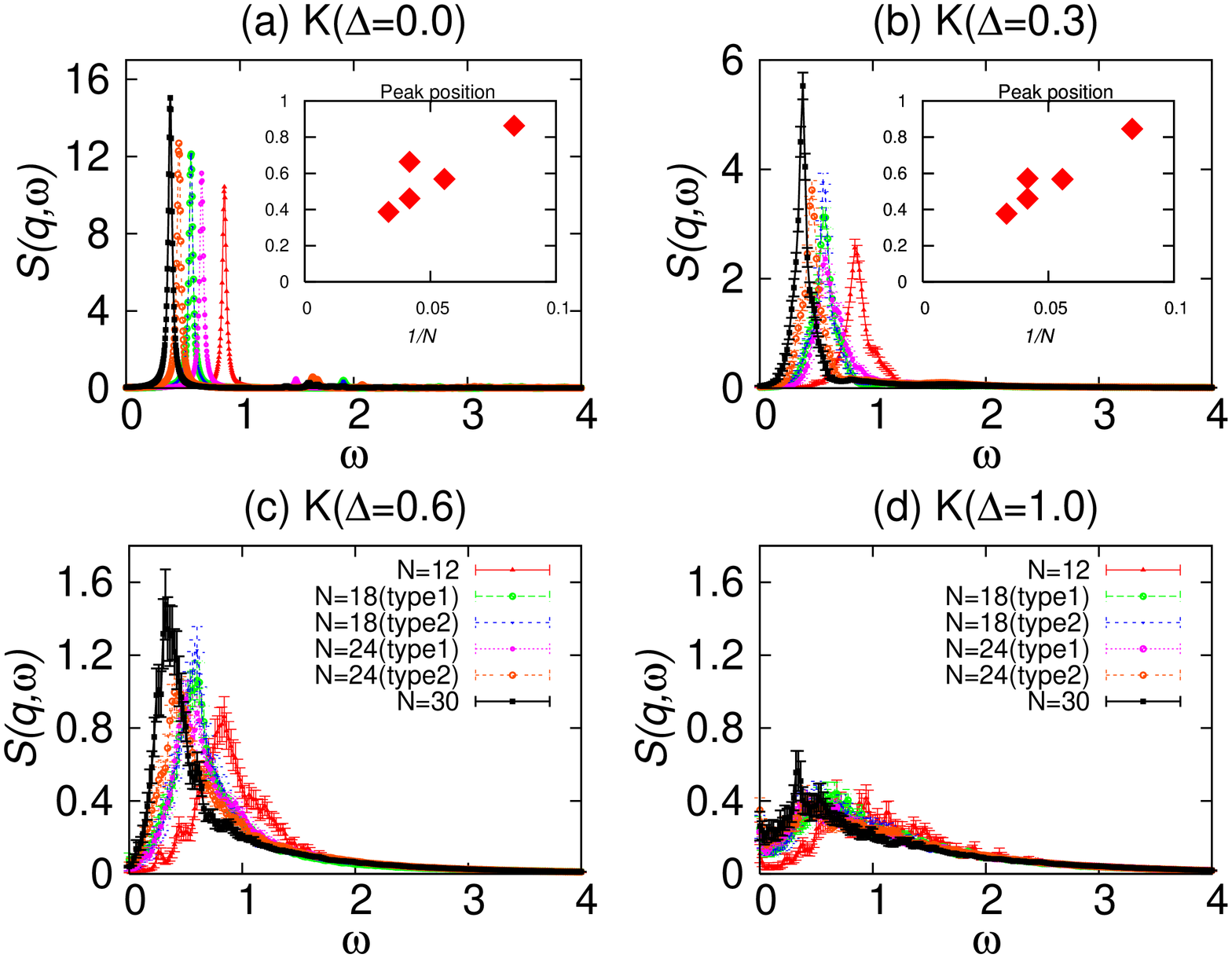}
 \caption{
(Color online) The $\omega$-dependence of the dynamical structure factors of the triangular-lattice Heisenberg antiferromagnet taken at the K point, ${\bf q}=(\pm 2\pi/3, \pm2\pi/\sqrt{3})$ and $(\pm 4\pi/3, 0)$, for  several values of the randomness $\Delta=0$ (a), 0.3 (b), 0.6 (c), and 1.0 (d). The critical randomness separating the AF state and the random-singlet state is $\Delta_{\rm c}\simeq 0.6$.  Note the difference in the ordinate scale between Figs.(a) and (b)-(d). In the inset of Figs.(a) and (b), the $\omega$ value of the dominant peak is plotted versus the inverse system size $1/N$.
}
\end{figure}

\begin{figure}[t]
 \includegraphics[width=8.5cm]{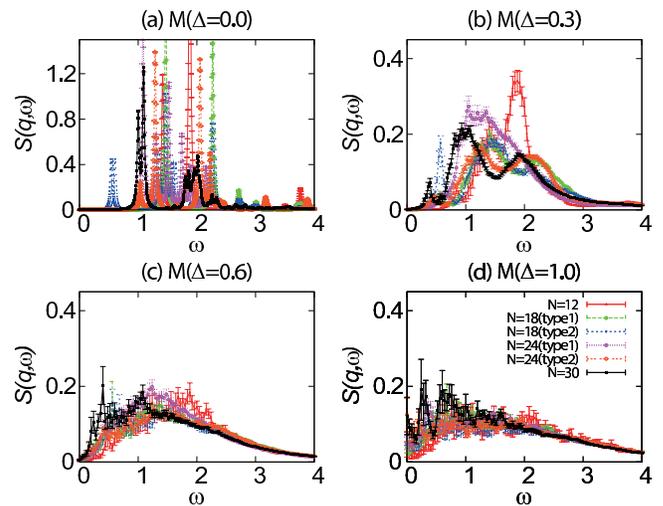}
 \caption{
(Color online)  The $\omega$-dependence of the dynamical structure factors of the triangular-lattice Heisenberg antiferromagnet taken at the M point, ${\bf q}=(0, \pm 2 \pi / \sqrt{3})$ and $(\pm \pi, \pm \pi / \sqrt{3})$,  for  several values of the randomness $\Delta=0$ (a), 0.3 (b), 0.6 (c), and 1.0 (d). The critical randomness separating the AF state and the random-singlet state is estimated to be $\Delta_{\rm c}\simeq 0.6$.  Note the difference in the ordinate scale between  Figs.(a) and (b)-(d), and between this figure and Fig.3.
}
\end{figure}

\section{\label{sec:Kagome lattice}Results II: the kagome lattice}

 Next, we deal with the random-bond $S=1/2$ AF Heisenberg model on the kagome lattice whose Hamiltonian is given by eq.~(1). We first examine the temperature dependence of the two representative types of the rescaled sublattice magnetization per spin, $\tilde m_{\rm s}$, each associated with the $q=0$ and the ${\sqrt{3} \times \sqrt{3}}$ structures. As in the triangular case, it is normalized to give unity for the classical, perfectly ordered $q=0$ or ${\sqrt{3} \times \sqrt{3}}$ structure, while the auto-correlated part is subtracted. The temperature dependence of the computed $\tilde m_{\rm s}$ is shown in Fig.5 for $N=18$ for several values of the randomness $\Delta$.

 Numerically, it has been established that the regular model exhibits neither the $q=0$ nor the $\sqrt 3\times \sqrt 3$ LRO even at $T=0$\cite{Lecheminant}. These AF orders are not realized in the random model, either\cite{Kawamura}. Yet, the temperature and the size dependence of $\tilde m_{\rm s}$ is expected to provide us useful information about the associated AF short-range order. 

 For the regular model, while the spin-wave-type $1/S$ expansion suggested the possible dominance of the $\sqrt 3\times \sqrt 3$ order \cite{Chubukov,Chernyshev}, the recent numerical results from the exact-diagonalization \cite{Lauchli2} and the DMRG \cite{Jiang} calculations suggested the dominance of the $q=0$ state. Though our present data for $N=30$ apparently suggest the dominance of the $\sqrt 3\times \sqrt 3$ short-range order, our maximum size $N=30$ is smaller than the ones of Ref.\cite{Lauchli2} and \cite{Jiang}, {\it i.e.\/}, $N=36$ and $N=108$, repsectively, and might be subject to stronger finite-site effect. In any case, as the randomness gets stronger, the difference between the $q=0$ order and the $\sqrt 3\times \sqrt 3$ order tends to be negligible.

 To get further information about the spin correlations of the model, we compute the $T=0$ static spin structure factor, and the results for $N=30$ are shown in Fig.6 as an intensity plot in the ($q_x,q_y$)-plane for the randomness of $\Delta =0$ (a), 0.4 (b), 0.7 (c), and 1.0 (d). The length unit here is taken to be the nearest-neighbor distance of the original kagome lattice.

 In the regular case shown in Fig.6(a), the intensity appears primarily along the zone boundary of the extended BZ, consistently with the behavior reported in Refs.\cite{Lauchli2, Jiang}. We note that the recent numerical studies for larger system of $N=36$ reveal additional small peaks at the wavevectors corresponding to the $q=0$ state, ${\bf q}=(0, \pm 2 \pi/\sqrt{3})$ and $(\pm \pi, \pm \pi/\sqrt{3})$, which are located along the zone boundary of the extended BZ \cite{Lauchli2, Jiang}, although this peak is not clearly discernible in our present data of $N=30$.

 As can be seen from Figs.6(a)-(d), qualitative features of $S({\bf q})$ do not change much even when the randomness is introduced up to $\Delta=1$, except that the ridge-like intensity along the extended BZ boundary is somewhat broadend.

\begin{figure}[t]
 \includegraphics[width=8.5cm]{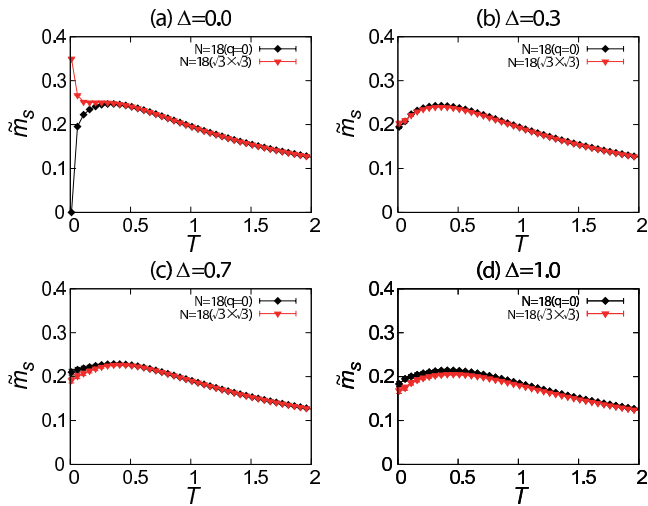}
 \caption{(Color online) The temperature dependence of the rescaled sublattice magnetization per spin, $\tilde m_{\rm s}$, associated with the $q=0$ order and the $\sqrt{3} \times \sqrt{3}$ order of the kagome-lattice Heisenberg antiferromagnet for several values of the randomness $\Delta=0$ (a), 0.4 (b), 0.7 (c), and 1.0 (d).
}
\end{figure}

\begin{figure}[t]
 \includegraphics[width=8.5cm]{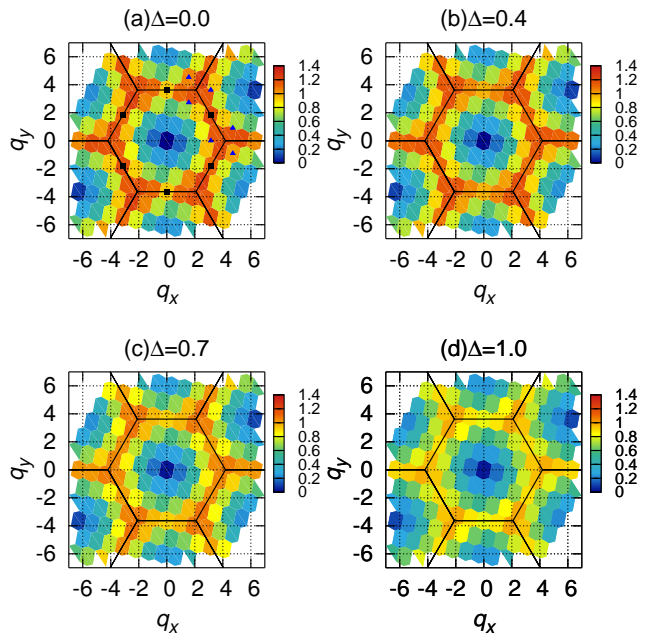}
 \caption{
(Color online) Intensity plots of the static spin structure factors $S({\bf q})$ of the kagome-lattice Heisenberg antiferromagnet in the wavevector ($q_x,q_y$) plane for several values of the randomness $\Delta=0$ (a), 0.4 (b), 0.7 (c), and 1.0 (d). The lattice constant $a=1$ is the nearest-neighbor distance of the kagome lattice. The system size is $N=30$. The solid black line depicts the zone boundary of the extended BZ. The black square in Fig.(a) represents the $\Gamma$ point, while the blue triangle represents the M point.
}
\end{figure}

\begin{figure}[t]
 \includegraphics[width=8.5cm]{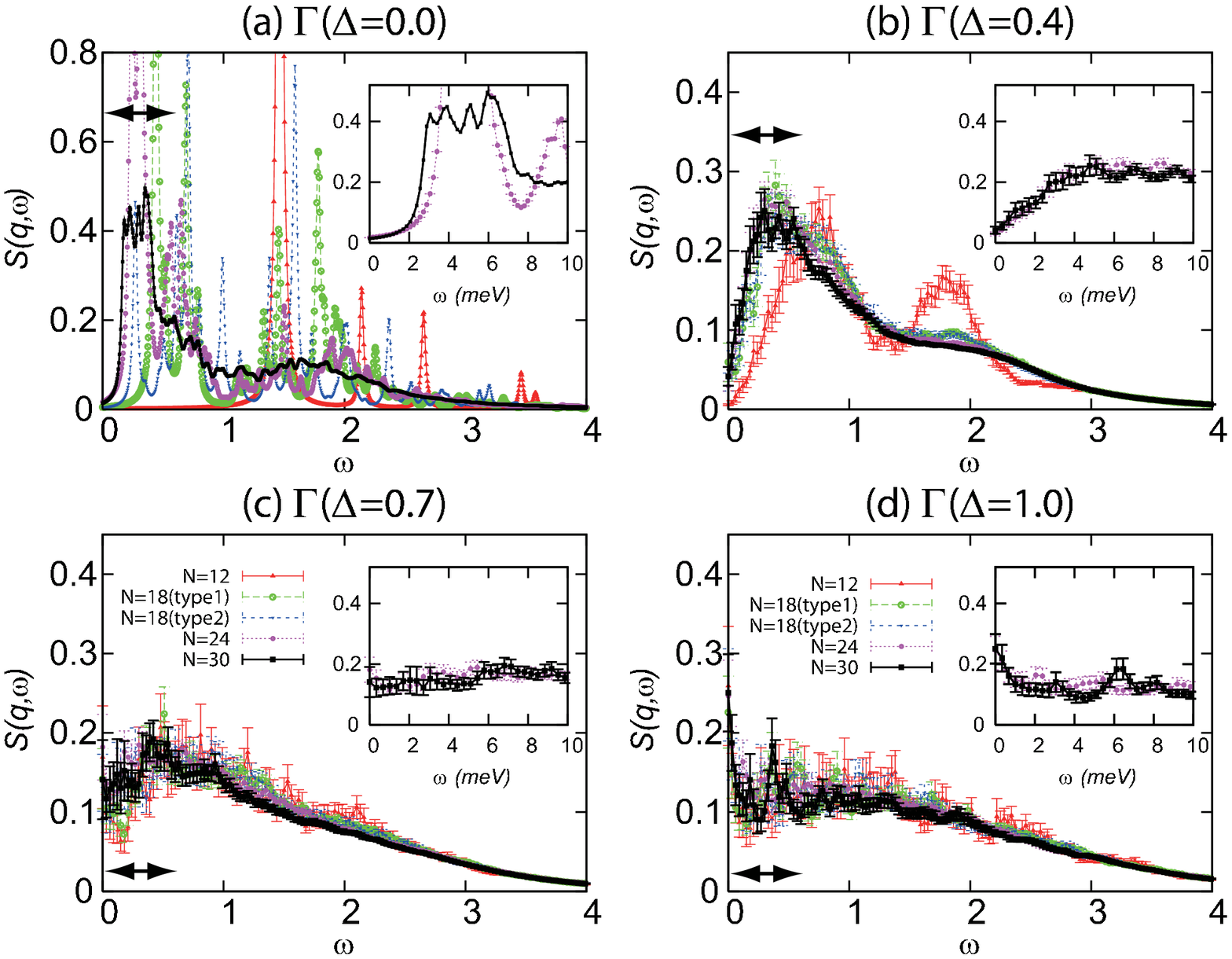}
 \caption{
(Color online)  The $\omega$-dependence of the dynamical spin structure factor $S({\bf q},\omega)$ of the kagome-lattice Heisenberg antiferromagnet taken at the $\Gamma$ point, ${\bf q}=(0, \pm 2 \pi/\sqrt{3})$ and $(\pm \pi, \pm \pi/\sqrt{3})$, for several values of the randomness $\Delta=0$ (a), 0.3 (b), 0.6 (c), and 1.0 (d). The critical randomness separating the randomness-irrelevant QSL stae and the random-singlet state is estimated to be $\Delta_{\rm c}\simeq 0.4$. The insets are magnified views of the low-$\omega$ region in units of meV, where $J=17$ meV is assumed with herbertsmithite in mind. Note the difference in the ordinate scale between  Figs.(a) and (b)-(d).
}
\end{figure}

\begin{figure}[t]
 \includegraphics[width=8.5cm]{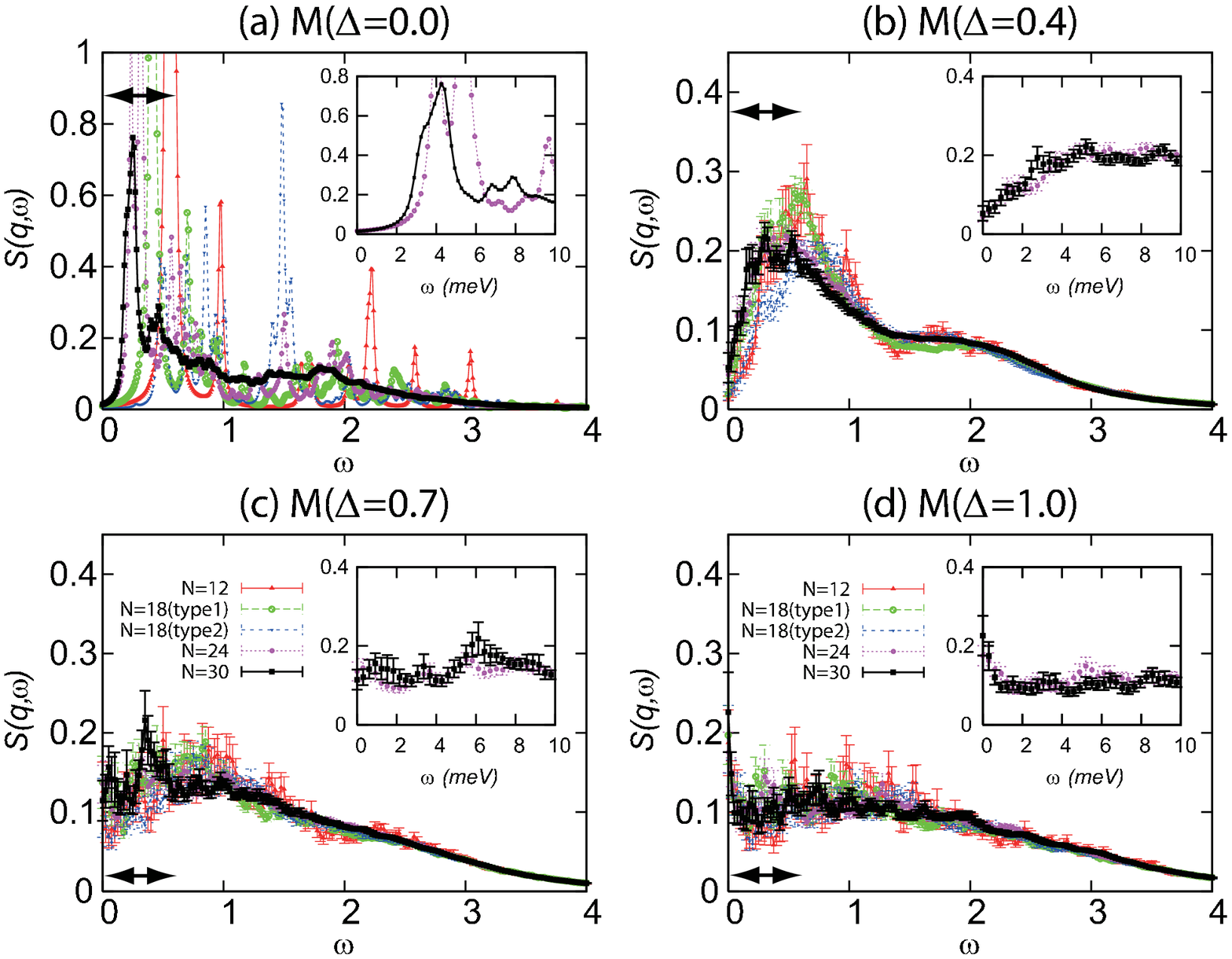}
 \caption{
(Color online)  The $\omega$-dependence of the dynamical spin structure factor $S({\bf q},\omega)$ of the kagome-lattice Heisenberg antiferromagnet taken at the M point, ${\bf q}=(0, \pm \pi /\sqrt{3})$ and $(\pm \pi/2, \pm \pi/2 \sqrt{3})$, for several values of the randomness $\Delta=0$ (a), 0.3 (b), 0.6 (c), and 1.0 (d). The critical randomness separating the randomness-irrelevant QSL stae and the random-singlet state is estimated to be $\Delta_{\rm c}\simeq 0.4$.  The insets are magnified views of the low-$\omega$ region in units of meV, where $J=17$ meV is assumed with herbertsmithite in mind. Note the difference in the ordinate scale between  Figs.(a) and (b)-(d).
}
\end{figure}

 We also compute the dynamical spin structure factor $S({\bf q},\omega)$ at the two representative wavevectors, {\it i.e.\/}, the so-called $\Gamma$ point, ${\bf q}=(0, \pm 2 \pi/\sqrt{3})$ and $(\pm \pi, \pm \pi/\sqrt{3})$, and the M point, ${\bf q}=(0, \pm \pi /\sqrt{3})$ and $(\pm \pi/2, \pm \pi/2 \sqrt{3})$. The computed $\omega$-dependence of  $S({\bf q},\omega)$ is shown for several values of the randomness $\Delta$ in Fig.7 for the $\Gamma$ point, and in Fig.8 for the M point. The $\omega$-dependence of the dynamical spin structure factor shows some differences between in the randomness-irrelevant QSL state at $\Delta < \Delta_{\rm c}$ and in the randomness-relevant QSL state at $\Delta > \Delta_{\rm c}$. In contrast to the case of the triangular model, $S({\bf q},\omega)$ exhibits a rather broad distribution even in the regular case, yet with some pronounced peaks remaining both at the $\Gamma$- and the M points as shown in Figs.7(a) and 8(a). Although our data of $N\leq 30$ are still subject to considerable finite-size effects, L\"auchli {\it et al.} reported that $S({\bf q},\omega)$ for $N=24$ and for larger $N=36$ came close, suggesting that $S({\bf q},\omega)$ for $N=30$ were not far from that of the bulk system. The issue of a small non-zero gap exists or not in the regular systems is beyond the capability of our present calculation.

 For larger $\Delta>\Delta_{\rm c}\simeq 0.4$ corresponding to the random-singlet state, by contrast, $S({\bf q},\omega)$ exhibits both at the $\Gamma$ and M points a nealy flat distribution in the wide $\omega$-range of, say, $\omega \lesssim 1.5$ and extends to higher $\omega$, with a nonzero intensity growing at $\omega=0$. This domonstrates the gapless nature of excitations in the random-singlet state both for the $\Gamma$- and M points. In the case of the strongest randomness $\Delta=1.0$, even an $\omega=0$ peak appears at both $\Gamma$ and M points. As can be seen from Fig.7(d) and 8(d), the overall behavior of $S({\bf q},\omega)$ are very much similar at the $\Gamma$ and the M points in the random-singlet state.

 In the random-singlet state, $S({\bf q},\omega)$ exhibits a tail in $\omega$ in the higher-$\omega$ range both at the $\Gamma$ and the M points. The asymptotic $\omega$-dependence of this tail is found to be exponential, $\approx \exp [-\omega/\omega_0]$, with a characteristic energy scale $\omega_0$. In the kagome model, such an exponential tail is also realized even in the regular system. Estimates of $\omega_0$ yields a value around 1.5$\sim $2, slightly increasing with increasing the randomess $\Delta$.

 The behavior of $S({\bf q},\omega)$ in the random-singlet state of the kagome model is rather similar to the one of the triangular model shown in Figs. 3 and 4. One difference might be that $S({\bf q},\omega)$ in the low-energy region is even more flatter in the kagome model, and the $\omega=0$ peak is more eminent.

 Anyway, the gapless behavior with an almost flat distribution extending to higher $\omega$ irrespective of the $q$-value is a common feature of the observed dynamical structure factor of both the triangular and the kagome models for larger $\Delta$, and might be regarded as a characteristic of the random-singlet state.

 In order to make comparison with the recent inelastic neutron-scattering data on a single-crystal of herbertsmithite\cite{Han}, we provide in the insets of Figs.~7 and 8 their low-energy part of the experimental data, which corresponds to the range indicated by the arrow in the main panel. Since the experimental data now available are limited to the low-energy region of $\omega\lesssim 10$meV, we show in the inset exactly this range, {\it i.e.\/}, $\omega < 10$ meV, by using an experimental estimate of $J\simeq 17$meV. This estimate of the exchange coupling $J$ of hermertsmithite is made via neutron-scattering measurements on powder samples\cite{Helton}. On comparison with the corresponding experimental data (Fig.2 of Ref.\cite{Han}), one finds a nice agreement especially for the strong randomness $\Delta=1$ shown in Fig.(d), including the features of (1) a plateau-like behavior observed for $2 \lesssim {\omega} \lesssim 10$ meV, (2) an $\omega=0$ peak observed at $\omega \lesssim 1.5$ meV without any gap, and (3) the insensitivity of the overall behavior of $S({\bf q},\omega)$ on the wavevector ${\bf q}$. We emphasize that the data of the random model shown in Fig.(d) appears to resemble the experimental one much more than that of the non-random system of Fig.(a). This observation certainly lends support to the view that the randomness is playing a significant role in the QSL state of herbertsmithite. 

\begin{figure}[t]
 \includegraphics[width=5.0cm,angle=270]{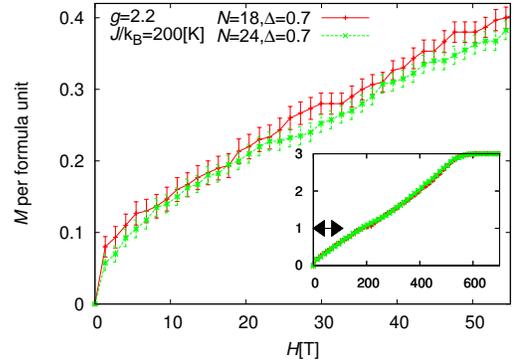}
 \caption{
(Color online) 
The $T=0$ magnetization curve for the randomness of $\Delta =0.7$ for $N=18$ and $24$,  in the wider field range (inset) and in the lower field range (main panel) corresponding to the region indicated by the arrow in the inset. With herbertsmithite in mind, the magnetization $M$ is normalized per formula unit of herbertsmithite, {\it i.e.\/}, the saturation value taken to be 3, while an applied field is given in units of tesla with assuming the experimental $g$-factor ($g\simeq 2.2$) and $J$-value ($J\simeq 200$K) of herbertsmithite \cite{Han2}.
}
\end{figure}

 Very good agreement with experiment is also found for the magnetization curve. In Fig.~9, we show the ground-state magnetization curve for the randomness $\Delta=0.7$ for $N=18$ and $24$, where the magnetization $M$ is normalized per formula unit of herbertsmithite, {\it i.e.\/}, the saturation value taken to be 3, while an applied field is given in units of tesla with assuming the experimental $g$-factor ($g\simeq 2.2$) and $J$-value ($J\simeq 200$K) of herbertsmithite \cite{Han2}. Then, the computed magnetization curve of Fig.9 indeed exhibits an almost quantitative agreement with the recent experimental data given in Fig.~2 of Ref.~\cite{Han2} in the same units.

Two features might be noticed. One is the absence of any plateau-like anomaly in the magnetization curve. As shown in the inset, it exhibits a near linear behavior in an entire region of $H$ up to the near saturation, except for some `wavy' behavior occurring at intermediate fields (which becomes less visible for the stronger randomness of $\Delta=1$ \cite{Kawamura}). As reported Ref.~\cite{Kawamura}, we have found that the plateau-like anomaly tends to go away for $\Delta > \Delta_{\rm c}$, yielding a near linear behavior. The other notable feature of the magnetization curve might be an upper-convex gapless behavior observed at weaker fields near $H=0$. This enhanced behavior of the low-field magnetization is consistent with the Curie-like behavior of the susceptibility observed for stronger randomness \cite{Kawamura}, and is likely to be borne by the `free' or `almost free' spins inevitably generated in the random-single state.

\section{\label{sec:SUMMARY}Summary and discussion}

 We studied by means of an ED method the nature of spin correlations of the random-bond $S=1/2$ AF Heisenberg models on the triangular and the kagome lattices via the static and the dynamical spin structure factors. To hilight the possible importance of frustration, we also made a comparative calculation for the unfrustrated random-bond $S=1/2$ AF Heisenberg model on the square lattice.

 Both the triangular and the kagome models exhibit the randomness-induced QSL behavior when the randomness exceeds a critical value as observed in previous studies, while the unfrustrated square model persistently exhibits the AF LRO up to the maximal randomness without showing the QSL behavior. This demonstrates that the frustration is certainly playing a role in stabilizing the random-singlet state.

 The random-singlet states in the triangular and the kagome models have some similarities, but also some differences. The random singlet state of the both models exhibit gapless behaviors, dependent on the wavevector ${\bf q}$ only weakly, while the dynamical spin structure factor $S({\bf q},\omega)$ exhibit a broad distribution in $\omega$ extending to higher $\omega$ with an exponentail tail. Especially in the strongly random kagome model, $S({\bf q},\omega)$ hardly depends on ${\bf q}$ and exhibits an almost flat distribution for a wide range of $\omega$ with a $\omega=0$ peak.

 As discussed in \S IV, our results for the dynamical spin structure factor for the strongly random kagome model compares quite favorably with the recent inelastic neutron-scattering data on the kagome herbersmithite, including (1) a plateau-like behavior observed for $2 \lesssim {\omega} \lesssim 10$ meV, (2) an $\omega=0$ peak observed $\omega \lesssim 1.5$ meV without any gap, and (3) the insensitivity on the wavevector ${\bf q}$. Since the present experimental data are limited to the low energy range $\omega\lesssim 10$ meV, it would be interesting to perform further experiments in the higher energy range of $\omega \gtrsim 10$ meV to make further comparison. In addition, the computed magnetization curve is found to exhibit a good, almost quantitative agreement with the recent experimental data on herbertsmithite.

In making a truly quantitative comparison with the experimental data on herbertsmithite, ones needs to examine several effects not considered in the present model, {\it i.e.\/}, the effect of the Dzyaloshinskii-Moriya interaction and the triangular layer between the kagome layers, {\it etc\/}. How the results depend or do not depend on the particular form of the randomness needs further clarification. In particular, the possible effect of the dilution-type randomness pointed out experimentally in Refs.\cite{Vries,Mendelse,Freedman} and studied theoretically in Ref.\cite{Singh3} might be incorporated.

 An important open theoretical question in the kagome model might be the distinction and the true relation between the observed two types of QSL-like states, {\it i.e.\/}, the randomness-irrelevant QSL state realized at $\Delta<\Delta_{\rm c}$ and  the randomness-relevant QSL state (random-singlet state) realized at $\Delta>\Delta_{\rm c}$. One intuitive view of this transition (or crossover) occurring at $\Delta\simeq \Delta_{\rm c}$ might be that singlet states tend to exhibit an Anderson localization there. The enhanced flat feature of $S({\bf q}, \omega)$ observed in the random singlet state at $\Delta>\Delta_{\rm c}$ is certainly consistent with such a picture. In this view, the randomness-irrelevant QSL state is an extended RVB-like state, while the randomness-relevant QSL state is an Anderson-localized state of singlets. Whether this naive picture captures correct physics behind the two types of QSL-like states needs further clarification.

 In contrast to the kagome case, the neutron-scattering data are not presently availabe for triangular organic salts due to the oganic nature of the material. Another route to the random-singlet state of $S=1/2$ triangular AFs might be an insulating mixed crystal Cs$_2$Cu(Br$_{1-x}$Cl$_x$)$_4$. Experiments suggest that this compound might exhibit a non-magnetic ground state in the range $x>0.17$ \cite{OnoTanaka}, at which the system possesses a considerable amount of randomness associated with the random arrangement of Cl$^-$ and Br$^-$. Interestingly, in this QSL-like regime, the gapless behavior including the $T$-linear low-temperature specific heat is observed \cite{Ono}. It might be interesting to perform neutron-scattering measurements on this compound in its QSL regime to make a comparison with our present data. 

The naive picture of the random singlet state, which is observed in common with the triangular and the kagome models, might be that tightly bound spin singlets are preferentially formed at stronger $J_{ij}$ bonds, leaving loosely bound spin singlets or nearly free spins at weaker  $J_{ij}$ bonds. Of course, such simple assignment of singlets to randomly distributed $J_{ij}$ bonds immediately meets contradiction or `frustration', revealing that the singlet formation in the spatially random environment is not a trivial matter at all. A subtle balance between the potential energy due to the local energy gain arising from the nonuniform $J_{ij}$ and the kinetic energy arising from the resonance of the local singlet states should determine the true ground state. In fact, in both cases of the triangular and the kagome models, we have observed that, although there generally exists a tendency that the strong singlet with larger negative ${\bf S}_i\cdot {\bf S}_j$-value tends to be formed at strong bonds with large $J_{ij}$-value, this tendency is quite often violated in the sense that strong local singlet is sometimes formed at weaker bonds, or the singlet formation remains weak even at stronger bonds.

 In summary, we investigated the nature of spin correlations in the randomness-induced QSL-like state, the random-singlet state, of the random-bond $S=1/2$ AF Heisneberg model on the triangular and the kagome lattices by computing their static and dynamical spin structure factors by means of the exact diagonalization method. Gapless behaviors accompnied by the broad distribution extending to higher energy, dependent on the wavevector only weakly, is observed in the dynamical spin structure factor $S({\bf q},\omega)$ in the random-singlet states of the both models in common. Especially in the kagome case with strong randomness, $S({\bf q},\omega)$ hardly dependeds on the wavevector ${\bf q}$ and exhibits an almost flat ditribution in a wide range of $\omega$, accompanied by the $\omega=0$ peak. These features agree with the recent inelastic-neutron scattering data on a single-crystal hermertsmithite semi-quantitatively, suggesting that the QSL state observed in herbertsmithite is indeed the random-induced QSL state, {\it i.e.\/}, the random-singlet state.

\section*{Acknowledgement}
The authors are thankful to Prof. K.~Kanoda, Prof. H.~Tanaka, Prof. I.~Ono, Prof. C.~Hotta, Prof. T.~Sakai and Dr.~K.~Hiji for useful discussion. Prof. H.~Tanaka pointed the possible importance of the Jahn-Teller distortion assiciated with the Cu$^{2+}$ in the triangular layer in assigning the quenched randomness in the exchange interaction of herbertsmithite. The ED calculation was performed by use of TITPACK Ver. 2. The QMC calculation was carried out using the ALPS QMC application\cite{ALPS}. They are thankful to ISSP, the University of Tokyo for providing us with CPU time. This study is supported by a Grant-in-Aid for Scientific Research No. 25247064. One of the authors (T.S.) acknowledges the financial support from the Motizuki Fund of Yukawa Memorial Foundation.

\appendix

\section{\label{sec:Appendix A}}

In this appendix, as a typical model of unfrustrated $S=1/2$ random systems, we deal with the random-bond $S=1/2$ AF Heisenberg mdel on the square lattice. The Hamiltonian is given by eq.~(1), the random interaction $J_{ij}$ obeying the uniform distribution characterized by the randomness parameter $\Delta$, commonly with the triangular and the kagome cases. The ED study on the bond-randomness effect in the unfrustrated square model have already been made in Ref.\cite{Nicolas}. It was shown there that, in the square-lattice case, the subalttice magnetization survived against the bond randomness. In this appendix, we show not only the ground-state sublattice magnetization but also the temperature dependence of the sublattice magnetization, together with the static and the dynamical ground-state spin structure factors, in order to make full comparison with the corresponding data for the frustrated systems.

 The lattice size is $N=8, 10, 16, 18, 20, 24$ and 32 with periodic boundary conditions in all directions. The sample average is performed for 300 ($N=8, 10$), 200 ($N=16, 18$), 100 ($N=20, 24$), and 5 ($N=32$) independent bond realizations in calculating the sublattice magnetization, and for 100 ($N<32$) and 20 ($N=32$) independent bond realizations in calculating the static and the dynamical spin structure factors.

\begin{figure}[ht]
 \includegraphics[width=5cm, angle=270]{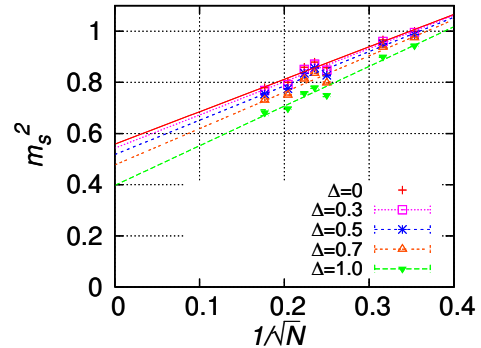}
 \caption{
(Color online) The rescaled $T=0$ squared sublattice magnetization per spin, $m_s^2$, of the square-lattice Heisenberg antiferromagnet plotted versus $1/\sqrt{N}$ ($N$ the lattice size) for several values of the randomness $\Delta$.
}
\end{figure}

\begin{figure}[ht]
 \includegraphics[width=6cm]{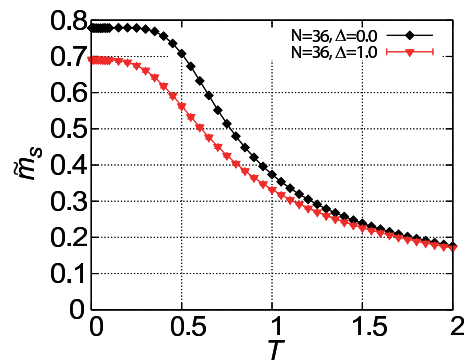}
 \caption{(Color online) The temperature dependence of the rescaled sublattice magnetization per spin, $\tilde{m}_{\rm s}$, of the square-lattice Heisenberg antiferromagnet for the size $N=36$, and for the randomness $\Delta=0$ and $1.0$.
}
\end{figure}

\begin{figure}[ht]
 \includegraphics[width=8.5cm]{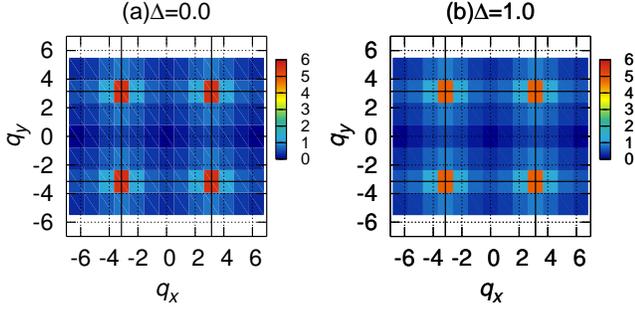}
 \caption{
(Color online) Intensity plots of the $T=0$ static spin structure factor $S({\bf q})$ of the square-lattice Heisenberg antiferromagnet in the  wavevector ($q_x,q_y$)-plane for the randomness $\Delta=0$ (a), and 1 (b). The solid black line represents the zone boundary of 1st BZ of the square lattice.
}
\end{figure}

\begin{figure}[ht]
 \includegraphics[width=3.2cm, angle=270]{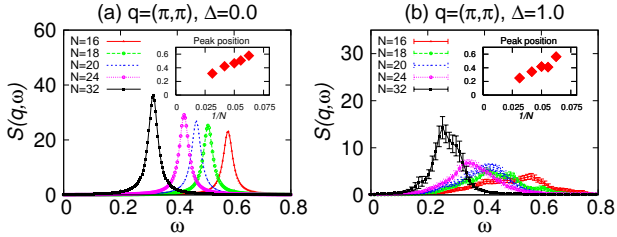}
 \caption{
(Color online) The $\omega$-dependence of the $T=0$ dynamical spin structure factor $S({\bf q},\omega)$ of the square-lattice antiferromagnet taken at the wavevector ${\bf q}={(\pi, \pi)}$ for the randomness $\Delta=0$ (a), and 1 (b).
}
\end{figure}

 The size dependence of the $T=0$ squared sublattice magnetization per spin, $m_{\rm s}^2$, is shown in Fig.10 for various randomness $\Delta$. The sublattice magnetization here is the one associated with the two-sublattice AF order. As can clearly be seen from the figure, $m_{\rm s}$ remains nonzero even in the the
rmodynamic limit for all values of $\Delta$, indicating that the AF LRO persists up to the maximal randomness of $\Delta=1$ as reported in ref.~[82]. This is in sharp contrast to the cases of the triangular and the kagome models where the AF LRO gives way to the random-singlet state for sufficiently strong randomness. Hence, not only the randomness but also the frustration plays a significant role in stabilizing the random-singlet state. In other words, all three elements, {\it i.e.\/}, the strong quantum fluctuation, the frustration, and the quenched randomness, conspire to realize the present QSL state, the random-singlet state. 

 In Fig.11, we show the temperature dependence of the rescaled sublattice magnetization per spin, $\tilde m_s$, for the regular $\Delta=0$ and for the maximally random $\Delta=1$ cases. The lattice size is $N=36$. The calculation here is made by use of the quantum Monte Carlo method\cite{ALPS}, which is possible due to the absence of frustration in the square-lattice model. In both the regular and the maximally random cases, $\tilde m_s$ increases monotonically with decreasing the temperature down to the lowest temperature studied, as can be seen from Fig.11. In particular, in contrast to the maximally random $\Delta=1$ case of the triangular model and the general case of the kagome model, the crossover behavior associated with a decrease of $\tilde m_s$ in the lower temperature regime is not observed in the square mdoel even in the maximally random case. This suggests that the crossover behavior observed in the triangular and the kagome models might be a characteristic of the formation of the QSL-like state including the random-singlet state.

 The $T=0$ static spin structure factor $S({\bf q})$ is shown in Fig.12 as an intensity plot in the wavevector plane. The lattice size is $N=24$. The computed $S({\bf q})$ exhibit rather sharp peaks at ${\bf q}=(\pm \pi, \pm \pi)$ corresponding to the AF LRO of the model both for $\Delta=0$ and $\Delta=1$. In fact, overall $S({\bf q})$ looks quie similar between the regular case of $\Delta=1$ (a) and the maximal random case of $\Delta=1$ (b). 

The $\omega$-dependence of the dynamical spin structure factor $S({\bf q},\omega)$ at a wavevector ${\bf q}=(\pi, \pi)$ corresponding to the AF order is shown in Fig.13 for the randomness $\Delta=0$ (a), and for $\Delta=1$ (b). The dominant peak is a single magnon excitation, the peak location tending to zero toward the thermodynamic limit as shown in the insets. As in the static case, $S({\bf q},\omega)$ looks qualitatively similar between the regular and the maximally random cases, while the magnon peak becomes broader in the random case.

\section{\label{sec:Appendix B}}

In this appendix, we show the lattice shapes used in our ED calculation for various lattice sizes $N$. The cases of the triangular lattice, the kagome lattice and the square lattice are given in Figs.14, 15 and 16, respectively. In all cases, periodic boundary conditions are applied in all directions.

\begin{figure}[h]
 \includegraphics[width=7.2cm]{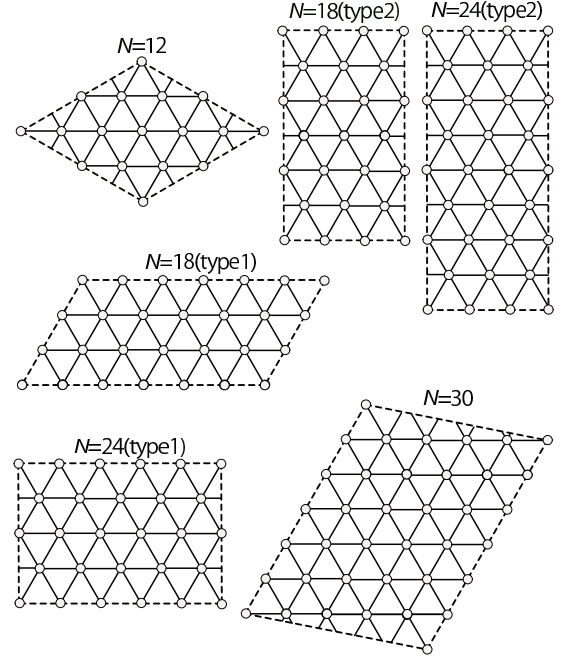}
 \caption{(Color online) The lattice shapes used in the exact diagonalization calculation of the triangular-lattice model for various $N$. Periodic boundary conditions are applied in all directions.
}
\end{figure}

\begin{figure}[ht]
 \includegraphics[width=7.2cm]{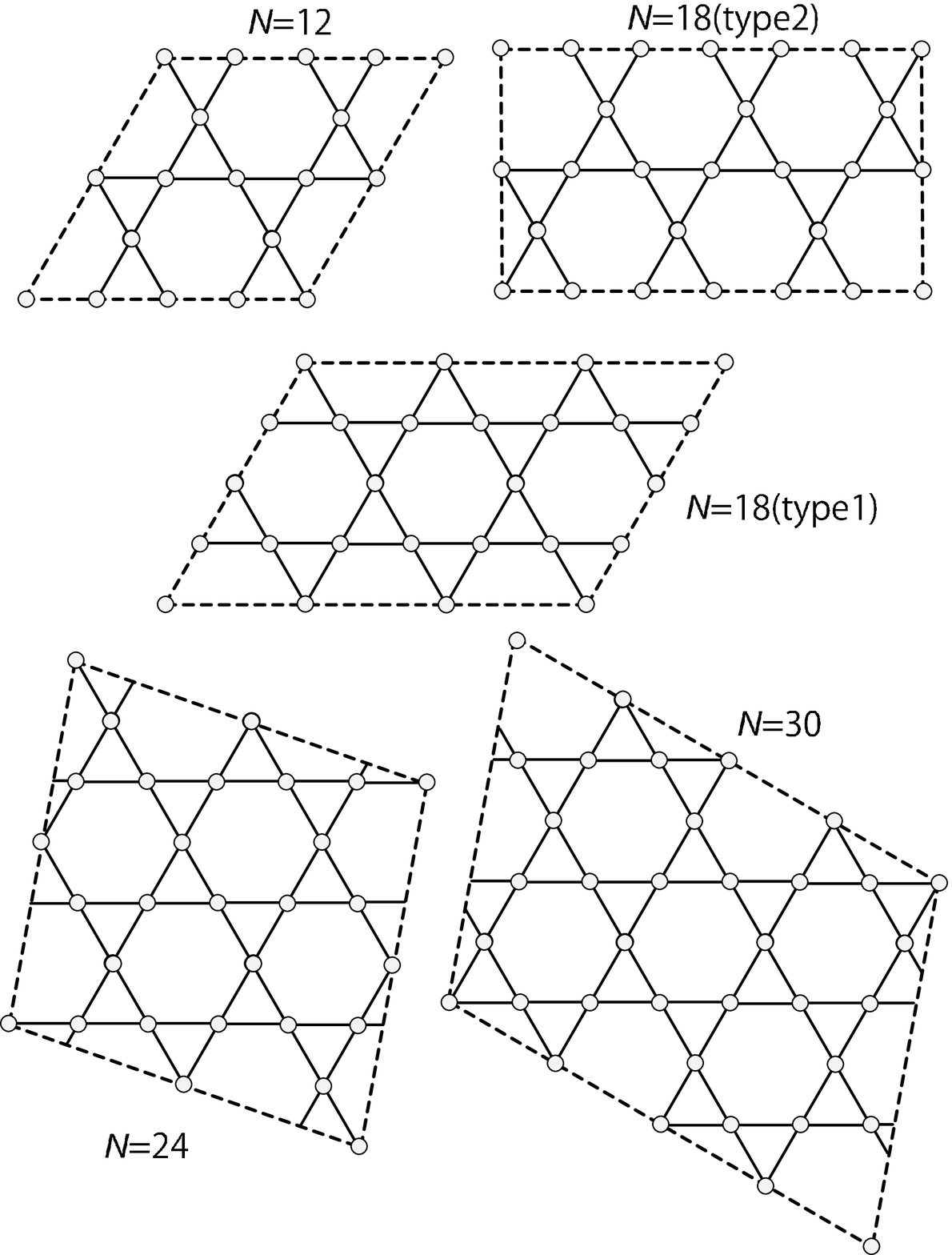}
 \caption{(Color online) The lattice shapes used in the exact diagonalization calculation of the kagome-lattice model for various $N$. Periodic boundary conditions are applied in all directions.
}
\end{figure}

\begin{figure}[ht]
 \includegraphics[width=7.2cm]{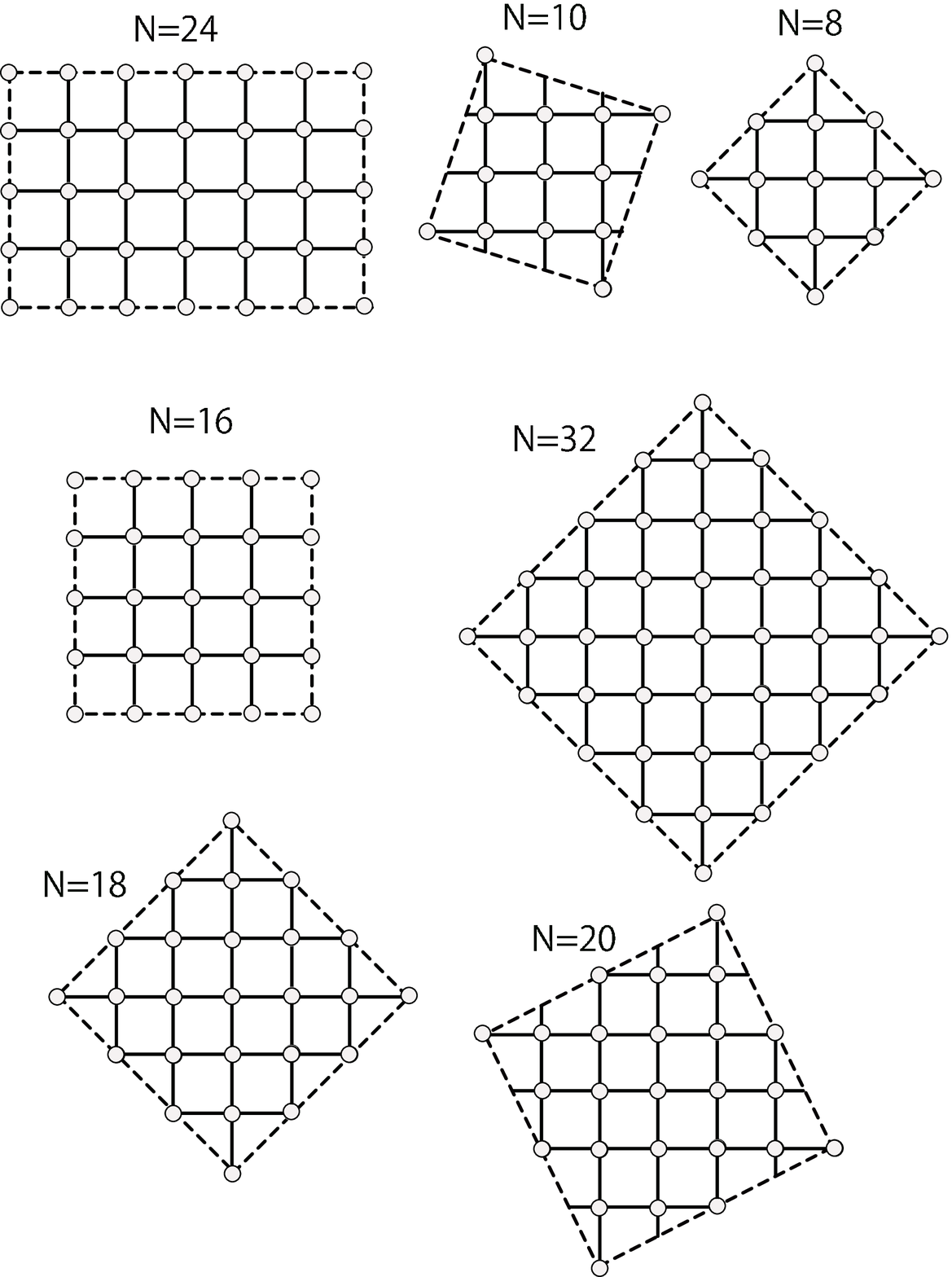}
 \caption{(Color online) The lattice shapes used in the exact diagonalization calculation of the sqaure-lattice model for various $N$. Periodic boundary conditions are applied in all directions.
}
\end{figure}

\end{document}